\documentclass[11pt]{article}
\usepackage{amsmath,amssymb,color,graphics,epsfig,cite}

\usepackage{amsfonts}

\newcommand{\hoch}[1]{$\, ^{#1}$}

\newcommand{\be}{\begin{equation}}
\newcommand{\ee}{\end{equation}}
\newcommand{\bea}{\setlength\arraycolsep{2pt} \begin{eqnarray}}
\newcommand{\eea}{\end{eqnarray}}
\newcommand{\nn}{\nonumber}

\def\ft#1#2{{\textstyle{\frac{\scriptstyle #1}{\scriptstyle #2} } }}
\def\fft#1#2{{\frac{#1}{#2}}}

\def\0{{\sst{(0)}}}
\def\1{{\sst{(1)}}}
\def\2{{\sst{(2)}}}
\def\3{{\sst{(3)}}}
\def\4{{\sst{(4)}}}
\def\5{{\sst{(5)}}}
\def\6{{\sst{(6)}}}
\def\7{{\sst{(7)}}}
\def\8{{\sst{(8)}}}
\def\sst#1{{\scriptscriptstyle #1}}

\begin{document}
\begin{flushright}
\hfill{USTC-ICTS/PCFT-21-33}
\end{flushright}

\begin{center}
{\Large {\bf Hidden Relations of Central Charges and OPEs\\ in Holographic CFT}}

\vspace{20pt}

{\large Yue-Zhou Li\hoch{1,2}, H. L\"u\hoch{1,3} and Liang Ma\hoch{1}}

\vspace{10pt}

{\it \hoch{1}Center for Joint Quantum Studies and Department of Physics,\\
School of Science, Tianjin University, Tianjin 300350, China
}

{\it \hoch{2}Department of Physics, McGill University, 3600 Rue University, Montr\'{e}al, QC Canada}

{\it \hoch{3}Peng Huanwu Center for Fundamental Theory, Hefei, Anhui 230026, China
}

\vspace{40pt}

\underline{ABSTRACT}
\end{center}

It is known that the $(a,c)$ central charges in four-dimensional CFTs are linear combinations of the three independent OPE coefficients of the stress-tensor three-point function. In this paper, we adopt the holographic approach using AdS gravity as an effect field theory and consider higher-order corrections up to and including the cubic Riemann tensor invariants. We derive the holographic central charges and OPE coefficients and show that they are invariant under the metric field redefinition.  We further discover a hidden relation among the OPE coefficients that two of them can be expressed in terms of the third using differential operators, which are the unit radial vector and the Laplacian of a four-dimensional hyperbolic space whose radial variable is an appropriate length parameter that is invariant under the field redefinition. Furthermore, we prove that the consequential relation $c=1/3 \ell_{\rm eff}\partial a/\partial\ell_{\rm eff}$ and its higher-dimensional generalization are valid for massless AdS gravity constructed from the most general Riemann tensor invariants.

\vfill{\footnotesize  liyuezhou@physics.mcgill.ca \ \ \ mrhonglu@gmail.com \ \ \ liangma@tju.edu.cn}


\thispagestyle{empty}
\pagebreak

\tableofcontents
\addtocontents{toc}{\protect\setcounter{tocdepth}{2}}

\newpage

\section{Introduction}

In a curved space, a conformal field theory (CFT) suffers from an anomaly that the trace of the stress tensor acquires a nonvanishing expectation value \cite{Duff:1977ay,Duff:1993wm}. In $d=4$, this conformal anomaly reads (we discard the Maxwell part)
\bea
\label{comformal anomaly}
\langle T^i_i \rangle=-\frac{a}{16\pi^2}E^{(4)}+\frac{c}{16\pi^2}I^{(4)},
\eea
where $E^{(4)}$ and $I^{(4)}$ represent the Euler density and the Weyl tensor squared of the geometric background respectively. Central charges $a$ and $c$ are important characteristics of the CFT, especially for the stress-tensor sector. Specifically, $a$-charge measures massless degrees of freedom and is the value of the $a$-function evaluated at the fixed point \cite{Zamolodchikov:1986gt,Cardy:1988cwa,Komargodski:2011vj}, and $c$-charge is proportional to the canonical normalization of stress-tensor two-point function \cite{Osborn:1993cr}, i.e., $C_T|_{d=4}=40/\pi^4 c$, where
\begin{eqnarray}
\langle T_{ij}(x)T_{kl}(y)\rangle=\frac{C_T\mathcal{I}_{ijkl}(x-y)}{(x-y)^{2d}}\,.
\end{eqnarray}
They encode the coefficient $\lambda_{TTT}$ of the operator product expansion (OPE) of the stress-tensor \cite{Erdmenger:1996yc}, can further probe the conformal collider physics \cite{Hofman:2008ar,Hofman:2016awc} and the averaged null energy condition (ANEC) \cite{Hartman:2016lgu}.

In $\mathcal{N}=4$ super Yang-Mills (SYM) theory, the two central charges are identical, but they are not in general the same, e.g., $\mathcal{N}=1,2$ supersymmetric theories \cite{Hofman:2008ar}. It is insightful to study the large $N$ CFTs with sparse gap ($\Delta_{\rm gap}\gg 1$), which are expected to have weakly-coupled local gravity duals in AdS \cite{Heemskerk:2009pn}. The central charges are encoded in the pure gravity sector, based on the holographic dictionary \cite{Maldacena:1997re,Gubser:1998bc,Witten:1998qj}. The causality analysis at the Regge limit in AdS gravity imposes CEMZ bound \cite{Camanho:2014apa}
\be
\fft{|a-c|}{c}\lesssim \fft{1}{\Delta_{\rm gap}^2}\,,
\ee
which was later obtained purely from the sparse CFTs \cite{Afkhami-Jeddi:2016ntf,Costa:2017twz,Afkhami-Jeddi:2017rmx} and extended to more general three-point functions (e.g.~\cite{Meltzer:2017rtf,Afkhami-Jeddi:2018own,Afkhami-Jeddi:2018apj}).

The above known relations among the central charges and also the OPEs are algebraic in nature. By studying ``massless'' higher-curvature gravities that have only massless graviton in AdS, a linear differential relation was proposed and it survived many checks \cite{Li:2018drw}. It is
\be
c=\frac{1}{3}\ell_{\mathrm{eff}}\frac{\partial a}{\partial\ell_{\mathrm{eff}}}\,,\label{relation 4d}
\ee
where $\ell_{\rm eff}$ is the (effective) radius of the AdS vacuum. In this paper, we consider pure gravity of the most general Riemann tensor invariants and derive the explicit general holographic formulae for the $(a,c)$ charges that enable us to prove this relation.

AdS gravity that corresponds to the large-$N$ and sparse CFT can also be approached as an effective field theory (EFT) and higher-order curvature invariants arise order-by-order perturbatively.  In this EFT approach, there are no massive modes within the appropriate cutoff and therefore it is necessarily to consider all possible such terms. However, we actually have less nontrivial higher-order terms since there is a redundancy that one can perform field redefinitions of the metric in terms of curvature tensors order-by-order without altering the physics. (This is referred to as the ``equivalence theorem'' \cite{Chisholm:1961tha}.) It is thus useful to express the
relation eq.~\eqref{relation 4d} in terms of the length parameter $\ell_{\rm inv}^{(0)}$ that is invariant under the field redefinition, namely
\begin{eqnarray}
c=\frac{1}{3}\ell_{\rm inv}^{(0)}\frac{\partial a}{\partial\ell_{\rm inv}^{(0)}}\,.\label{relation 4d1}
\end{eqnarray}
We consider AdS gravity up to and including the cubic order of the Riemann tensor polynomials. At this order, we can derive the three OPE coefficients $\lambda_{TTT}$ as well as the $(a,c)$ charges holographically. We verify that they are indeed invariant under the field redefinitions, and therefore so is eq.~\eqref{relation 4d1}. In doing so, we confirm the known result that the central charges are linear combinations of the OPE coefficients. Furthermore, we discover further hidden differential relations among the OPE coefficients by differential operators that live in the four-dimensional hyperbolic space whose radial coordinate is $\ell_{\rm inv}^{(0)}$.

The paper is organized as follows. In section 2, we consider pure gravity constructed from general Riemann tensor invariants.  With the assumption that the theory admits an AdS vacuum, we derive the explicit formulae for $(a,c)$ central charges and derive the differential relation \eqref{relation 4d} for massless gravities.  In section 3, we treat the higher-order curvature invariants perturbatively and consider the most general polynomial invariants up to and including the cubic order. We derive the proper $\ell_{\rm inv}^{(0)}$ that is invariant under the field redefinition and becomes $\ell_{\rm eff}$ when the theory reduces to massless gravity. We prove the relation \eqref{relation 4d1}. We find further differential relations among the OPE coefficients of the three-point function of the stress tensor.  We conclude the paper in section 4. In appendix A, we generalise our four-dimensional results to general dimensions. In appendix B, we present the field redefinitions in general dimensions and derive the invariant length parameters.  In appendix C, we study the OPE coefficients of the three-point function of the stress tensor.


\section{Holographic central charges    \label{section higher}}
\subsection{Generalties of higher derivative gravity }

For our purpose, we consider pure gravity theories constructed from general Riemann tensor invariants in $(d+1)$ spacetime dimensions:
\begin{eqnarray}
S=\frac{1}{16\pi G_N^{(d+1)}}\int d^{d+1}x\sqrt{-g}\mathcal{L}(R_{\mu\nu\rho\sigma},g^{\mu\nu})\,.\label{genlag0}
\end{eqnarray}
We assume that the theory admits an AdS vacuum of radius $\ell_{\mathrm{eff}}$.
It is instructive to introduce tensors $P_{\mu\nu\rho\sigma}$ and $C^{\mu\nu\rho\sigma}_{\alpha\beta\gamma\eta}$ \cite{Bueno:2016xff,Bueno:2016ypa}
\begin{eqnarray}
P_{\mu\nu\rho\sigma}=\frac{\partial\mathcal{L}}{\partial R^{\mu\nu\rho\sigma}}\Big|_{g^{\alpha\beta}},\qquad C^{\mu\nu\rho\sigma}_{\alpha\beta\gamma\eta}=\frac{\partial P^{\mu\nu\rho\sigma}}{\partial R^{\alpha\beta\gamma\eta}}\,.
\end{eqnarray}
When evaluated on the vacuum, the tensor structures are rigidly fixed
\begin{eqnarray}
 \bar{P}^{\mu\nu\rho\sigma}=&&2\xi_0 \bar{g}^{\mu[\rho}\bar{g}^{\sigma]\nu}\,,
\cr
\bar{C}^{\mu\nu\rho\sigma}_{\alpha\beta\gamma\eta}
=&&\xi_1 \Big(\delta^{[\mu}_{\alpha}\delta^{\nu]}_{\beta}\delta^{[\rho}_\gamma\delta^{\sigma]}_\eta+\delta^{[\rho}_\alpha \delta^{\sigma]}_\beta \delta^{[\mu}_\gamma \delta^{\nu]}_\eta \Big)+\xi_2 \Big((\bar{g}^{\mu\rho}\bar{g}^{\nu\sigma}-\bar{g}^{\mu\sigma}\bar{g}^{\nu\rho})(\bar{g}_{\alpha\gamma}\bar{g}_{\beta\eta}-\bar{g}_{\alpha\eta}\bar{g}_{\beta\gamma})  \Big)
\cr &&
+4\xi_3 \Big( \delta^{[\mu}_\tau \bar{g}^{\nu][\rho}\delta^{\sigma]}_{\varepsilon}\delta^{\tau}_{[\alpha}\bar{g}_{\beta][ \gamma}\delta^\varepsilon_{\eta]} \Big)\,.
\end{eqnarray}
In holography, as we will see shortly, the coefficients $\xi_{i=0\sim 3}$ contain all the information pertaining to the central charges and hence OPEs $\lambda_{TTT}$. Some useful identities were proved \cite{Bueno:2018yzo} and we quote them here
\begin{eqnarray}
\bar{\mathcal{L}}(\ell_{\mathrm{eff}})=&&-\fft{4d}{\ell^2_{\mathrm{eff}}}\xi_0\,,\qquad \bar{\mathcal{L}}'(\ell_{\mathrm{eff}})=\frac{4d(d+1)}{\ell^3_{\mathrm{eff}}}\xi_0\label{EOM AdS L}\,,\\
\mathfrak{h}(\ell_{\mathrm{eff}})\equiv &&\bar{\mathcal{L}}(\ell_{\mathrm{eff}})+
\frac{\ell_{\mathrm{eff}}}{d+1}\bar{\mathcal{L}}'(\ell_{\mathrm{eff}})=0\label{EOM AdS h}\,,\\
\mathfrak{h}'(\ell_{\mathrm{eff}})=&&\frac{4d(4\xi_1+4d\xi_3+4d(d+1)\xi_2+
(d-1)\xi_0\ell_{\mathrm{eff}}^2)}{\ell^5_{\mathrm{eff}}}\label{EOM AdS dh}\,.
\end{eqnarray}
Here, a prime denotes a derivative with respect to $\ell_{\rm eff}$, but with subtleties that should be clarified. To be precise, the Lagrangian in \eqref{genlag0} should be written as ${\cal L}={\cal L}(R_{\mu\nu\rho\sigma},g^{\mu\nu},\alpha)$ where $\alpha$ denotes all the coupling constants, including the bare cosmological constant $\Lambda_0$.  Off shell, these coupling constants are all independent of $\ell_{\rm eff}$, and they become related by the on-shell condition, namely the left equation in \eqref{EOM AdS L}.  The derivative is implemented off shell, but assuming that the metric is AdS of radius $\ell_{\rm eff}$, namely
\cite{Bueno:2016xff,Bueno:2016ypa}
\be
{\cal L}'(\ell_{\rm eff}) = \fft{\partial \bar {\cal L}}{\partial \bar R^{\mu\nu\rho\sigma}}
\fft{\partial \bar R^{\mu\nu\rho\sigma}}{\partial\ell_{\rm eff}}\,.\label{primedef}
\ee
We consider in general theories with a bare cosmological constant and we solve the on-shell condition by express the $\Lambda_0$ in terms of $\ell_{\rm eff}$ and other coupling constants.  It is clear that the derivative in \eqref{primedef} does not explicitly involve in $\Lambda_0$.  The rule for higher derivatives follows straightforwardly.

Gravity of Riemann invariants is in general a 4'th-order derivative theory and the spectrum contains not only the usual massless graviton $h^{(0)}_{ab}$, but also a massive spin-2 $h^{(M)}_{ab}$ and a massive scalar $h$. The effective Newton constant $G_{N\mathrm{eff}}^{(d+1)}$ and the masses of additional modes can be expressed in terms of $\xi_{i=0\sim3}$ \cite{Bueno:2016xff,Bueno:2016ypa}
\begin{eqnarray}
\frac{1}{G_{N\mathrm{eff}}^{(d+1)}}&=&
\frac{2}{G_{N}^{(d+1)}}(\xi_0+\frac{2}{\ell_{\mathrm{eff}}^2}(d-2)\xi_1)\,,\label{Gneff}\\
m^2_s&=&\frac{\xi_0\ell_{\rm eff}^2(d-1)+4(\xi_1+d(d+1)\xi_2+d\xi_3)}{(2\xi_1+(d+1)\xi_3+4d\xi_2)\ell_{\rm eff}^2}\,,\\
m^2_g&=&-\frac{\xi_0\ell_{\rm eff}^2+2(d-2)\xi_1}{(2\xi_1+\xi_3)\ell_{\rm eff}^2}\,.
\end{eqnarray}
It is important to note that first equation above is the coefficient of the kinetic term of the graviton and hence it does not explicitly depend on the bare cosmological constant $\Lambda_0$.
In our proof of eq.~\eqref{relation 4d}, we need to decouple both massive modes by requiring
\be
\label{massless condition}
2\xi_1+(d+1)\xi_3+4d\xi_2=0\,,\qquad
2\xi_1+\xi_3=0\,.
\ee
The resulting theory is phrased as massless gravity in \cite{Li:2018drw} that include quasi-topological gravities \cite{quasi0,quasi1,Bueno:2019ltp,Bueno:2019ycr}.

\subsection{General formula of holographic central charges}

The conformal anomaly in CFT$_4$ can be reconstructed from the AdS bulk by considering FG expansion \cite{Henningson:1998gx,Nojiri:1999mh}
\be
ds^2=\frac{\ell_{\mathrm{eff}}^2}{4\rho^2}d\rho^2+\frac{1}{\rho}g_{ij}dx^idx^j\,,\qquad g_{ij}=g_{0ij}+g_{1ij}\rho+g_{2ij}\rho^2+\cdots\,,
\ee
where $\rho\rightarrow0$ is the AdS boundary. We introduce the ultraviolet (UV) cutoff parameter $\epsilon>0$ and evaluate the boundary integrals at $\rho=\epsilon$. Correspondingly, the action can be expanded as
\be
S=\frac{1}{16\pi G_N^{(5)}}\int_{\rho\rightarrow\epsilon} d^5x\sqrt{-\hat g_5}\mathcal{L}=
\frac{1}{16\pi G_N^{(5)}}\int d^{4}x\sqrt{-g_0}\int_{\rho\rightarrow\epsilon} d\rho(\cdots+\frac{A}{\rho}+\cdots)\,,
\label{action}
\ee
where the coefficient $A$ of $\rho^{-1}$ is interpreted as conformal anomaly, because $\rho^{-1}$ gives rise to a logarithmic dependence of the UV cutoff
\bea
\label{anomaly}
S_{\mathrm{anomaly}}=-\int d^{4}x\sqrt{-g_0}\,\fft{1}{2}\langle T^i_i \rangle\log{\epsilon}\,,
\eea
which manifestly breaks the scaling invariance.

For convenience, we consider a reduced FG expansion: we explicitly construct an AdS with $S^2\times S^2$ boundary \cite{Li:2017txk}
\bea
ds^2=\frac{\ell_{\mathrm{eff}}^2}{4\rho^2}d\rho^2+\frac{1}{\rho}\left(f_1(\rho)d\Omega_1^2+f_2(\rho)d\Omega_2^2\right)\,,
\eea
and $(f_1,f_2)$ can be closely solved as the truncated FG expansion
\bea
&& f_1=f_{10}+f_{11}\rho+f_{12}\rho^2 + \cdots\,,\qquad f_2=f_{20}+f_{21}\rho+f_{22}\rho^2 + \cdots\,,
\eea
where
\bea
f_{11}=\frac{1}{6} (\gamma -2) \ell_{\mathrm{eff}} ^2\,,\quad f_{21}=\frac{(1-2 \gamma ) \ell_{\mathrm{eff}} ^2}{6 \gamma }\,.
\eea
The constants $f_{10}$ and $f_{20}$ simply represent the radii of the two spheres. We define the ratio $\gamma=f_{10}/f_{20}$, but take $f_{10}=f_{20}=1$. The Euler density gives the topological number of $S^2\times  S^2$, whilst the Weyl-squared on the boundary depends on $\gamma$:
\bea
\label{GBWeyl}
E^{(4)}=8,\qquad I^{(4)}=\frac{4}{3}\left(2+\gamma+\frac{1}{\gamma}\right).
\eea
We can then readily find
\bea
A=\frac{1+\gamma^2}{6\gamma}\ell_{\mathrm{eff}}^3\left(\xi_0+
\frac{4\xi_1}{\ell_{\mathrm{eff}}^2}\right)
-\frac{2}{3}\ell_{\mathrm{eff}}^3\left(\xi_0-
\frac{2\xi_1}{\ell_{\mathrm{eff}}^2}\right),
\eea
from which we can read off the central charges
\be
a=\fft{\pi\xi_0\ell_{\mathrm{eff}}^3}{4G_N^{(5)}},\qquad c=\fft{\pi\ell_{\mathrm{eff}}^3}{4G_N^{(5)}}\left(\xi_0+\frac{4\xi_1}{
\ell_{\mathrm{eff}}^2}\right)=\frac{\pi\ell_{\mathrm{eff}}^3}
{8G_{N\mathrm{eff}}^{(5)}}.
\ee
We can then show generally $C_T|_{d=4}=40/\pi^4 c$ in holography
since $C_{T}$ is necessarily proportional to $1/G_{N\mathrm{eff}}^{(5)}$. (For explicit low-lying examples, see e.g., \cite{Buchel:2009sk,Myers:2010jv}.) In Appendix \ref{app: relation in general dim}, we will generalize our results to arbitrary dimensions and prove generally that the $a$-charge above can be read off from the entanglement entropy of spherical entangling surfaces, demonstrated in some explicit low-lying examples \cite{Myers:2010xs,Myers:2010tj,Casini:2011kv}.

\subsection{Proof of the central charge relation}

The identities (\ref{EOM AdS dh}) and (\ref{Gneff}) imply that the relation between $G^{(5)}_{N \mathrm{eff}}$ and $\mathfrak{h}'(\ell_{\mathrm{eff}})$:
\bea
\frac{1}{G_{N\mathrm{eff}}^{(5)}}=\frac{\ell_{\mathrm{eff}}^3}{24G_N^{(5)}}
\left(\mathfrak{h}'(\ell_{\mathrm{eff}})+\frac{128}{\ell_{\mathrm{eff}}^2}
(\xi_1-10\xi_2-2\xi_3)\right).
\eea
The term $\mathfrak{h}'(\ell_{\mathrm{eff}})$ can also be obtained by a derivative of eq.~\eqref{EOM AdS h}, followed by substituting eq.~\eqref{EOM AdS L}. We have We have
\bea
\label{central charge relation}
c&=&\frac{\pi\ell^{3}_{\mathrm{eff}}}{12G_N^{(5)}}\left(3\xi_0+\ell_{\mathrm{eff}}
\frac{\partial \xi_0}{\partial\ell_{\mathrm{eff}}} +\frac{8}{\ell_{\mathrm{eff}}^2}(\xi_1-10\xi_2-2\xi_3) \right)\cr
&=&\frac{\pi}{12 G_N^{(5)}}\ell_{\mathrm{eff}}\frac{\partial(\xi_0\ell_{\mathrm{eff}}^{3})}
{\partial\ell_{\mathrm{eff}}}+\frac{2\pi}{3G_N^{(5)}}(\xi_1-10\xi_2-2\xi_3)\cr
&=&\frac{1}{3}\ell_{\mathrm{eff}}\frac{\partial a}{\partial\ell_{\mathrm{eff}}}+\frac{2\pi}{3G_N^{(5)}}(\xi_1-10\xi_2-2\xi_3).
\eea
The second term above vanishes after imposing the massless conditions \eqref{massless condition}. We therefore prove the universal relation \eqref{relation 4d} of holographic central charges for massless AdS gravities. It should be emphasized that since the massless conditions \eqref{massless condition} in general involve $\ell_{\rm eff}$, we should take the $\ell_{\rm eff}$ derivative of $a$ before imposing the massless conditions such that the relevant coupling constants are independent of $\ell_{\rm eff}$.

    In appendix A, we generalize the relation to arbitrary dimensions and the subtleties of the
$\ell_{\rm eff}$ derivative are further clarified with explicit examples.

\section{Holographic CFT as AdS EFT \label{section EFT}}

\subsection{General arguments}

In this section, we use the EFT approach to the weakly-coupled bulk gravity, where higher-order Riemann tensor polynomials are perturbative corrections to Einstein gravity with a bare negative cosmological constant
\be
\Lambda_0=-\fft{d(d-1)}{2\ell_0^2}\,.
\ee
For the theory to be valid below the energy scale $M\gg 1/\ell_0$, graviton is the only light particle and the massive states are all beyond $M$, i.e.,
\begin{eqnarray}
&& 2\xi_1+(d+1)\xi_3+4d\xi_2=\fft{\#_1}{M^2}+\fft{\#_2}{M^4\ell_0^2}+\cdots\,,
\cr &&
2\xi_1+\xi_3=\fft{\#_1'}{M^2}+\fft{\#_2'}{M^4\ell_0^2}+\cdots\,.\label{massless general}
\end{eqnarray}
Furthermore, we assume the following hierarchy \cite{Caron-Huot:2021enk}
\be
\fft{1}{G_N^{(d+1)}}\gg M^{d-1}\gg \fft{1}{\ell_0^{d-1}}\,,\quad \ell_0\sim \mathcal{O}(1)\,.
\ee
We consider only the tree-level gravitational physics, of the leading order in $\mathcal{O}(G_N^{(d+1)})$, which is dual to the large $C_T$ ($C_T \sim N^2$) limit of the boundary CFT. Schematically, we may write the effective Lagrangian as (we only consider parity-even gravity)
\bea
\mathcal{L}(R_{\mu\nu\rho\sigma},g^{\mu\nu})=R-2\Lambda_0+\sum_{n=2} \sum_i g_{n,i} [R^{(n)}_i]\,,\label{genlag}
\eea
where $g_{n,i}$'s are Wilson coefficients associated with higher derivative terms $[R^{(n)}_i]$ that denote $i$'th curvature invariant operators with $2n$ derivatives. In flat space, dimensional analysis indicates the scaling behavior $g_{n,i}\sim 1/M^{2(n-1)}$; however, it becomes more complicated in AdS. The subtlety arises from the fact that $g_{(n+1),i}$ can also enter the coupling $g_n$ under field redefinitions. Thus the most general power-counting for each  $g_{n,i}$ involves all $1/(M^{2(n-1-k)}\ell_0^{2k})$ terms with $k=0,1,2,\cdots$.

In this framework, $\#$ and $\#'$ are dimensionless quantities that depend on Wilson coefficients of higher derivative terms; however, they are not invariant under the field redefinition, and neither does $\ell_0$. It is always possible to find a field redefinition that brings eq.~\eqref{massless general} to eq.~\eqref{massless condition} \cite{Bueno:2019ltp}. As physical observables, central charges should be invariant under the field redefinitions. It is thus natural to expect the relation eq.~\eqref{relation 4d} should have a general form for AdS EFTs, where $\ell_{\rm eff}$ is replaced by a certain field-redefinition invariant quantity $\ell_{\rm inv}^{(0)}$ that coincides with $\ell_{\rm eff}$ for massless gravity. This leads to eq.~\eqref{relation 4d1}.

\subsection{Example: to the cubic order and OPE relations}

For a concrete example, we truncate the AdS EFT to the cubic order and show the field redefinition invariance of central charges and the validity of the relation eq.~\eqref{relation 4d1}. Cubic is also the minimum order that fully enumerates three-point structures of the stress tensor. (We shall return to this point later.) The complete set of quadratic and cubic operators is
\bea
\label{quadratic cubic theory}
 [R^{(2)}]=&& \{R^2,R_{\mu\nu}R^{\mu\nu},R_{\mu\nu\rho\sigma}R^{\mu\nu\rho\sigma}\}\,,
\cr
[R^{(3)}]=&& \{R^3,RR_{\mu\nu}R^{\mu\nu},R^\mu_\nu R^\nu_\rho R^\rho_\mu,
R_{\mu\nu\rho\sigma}R^{\mu\rho}R^{\nu\sigma},
RR_{\mu\nu\rho\sigma}R^{\mu\nu\rho\sigma},R^{\mu\nu}R_{\mu \rho\sigma\eta}R_\nu^{\ \rho\sigma\eta},
\cr && R^{\mu\nu}_{\ \ \rho\sigma}R^{\rho\sigma}_{\ \ \alpha\beta}R^{\alpha\beta}_{\ \ \mu\nu},R^{\mu\ \rho}_{\ \nu\ \sigma}R^{\nu\ \sigma}_{\ \alpha\ \beta}R^{\alpha\ \beta}_{\ \mu\ \rho},R_{\mu\nu}\Box R^{\mu\nu},R\Box R\}\,.
\eea
The contributions to central charges from this set were obtained in \cite{Li:2017txk} and we review them in Appendix \ref{app: field redefinition}.
To the cubic order, the Wilson coefficients can be expressed in terms of three sets of dimensionless parameters $(q_i,\tilde q_i, c_i)$:
\bea
\label{gnicoeff}
g_{2,i}&=&\fft{q_i}{M^2}+\fft{\tilde{q}_i}{M^4\ell_0^2}\,,\quad i=1,2,3\,,\nn\\
g_{3,i}&=&\fft{c_i}{M^4}\,,\quad\qquad\qquad i=1,2,\ldots,10\,.
\eea
We now show the rigidity of central charges under the general field redefinition:
\bea
g_{\mu\nu}\rightarrow d_0\, g_{\mu\nu}+\sum_id_{1i}[R^{(1)}_i]_{\mu\nu}+\sum_i d_{2i}[R^{(2)}_i]_{\mu\nu}\,,
\label{field redefinition}
\eea
where the $d_n$ coefficients are given explicitly in appendix \ref{app: field redefinition}, and
\bea
 [R^{(1)}]_{\mu\nu}=&&\{R_{\mu\nu}, R g_{\mu\nu}\}\,,
\cr
[R^{(2)}]_{\mu\nu}=&& \{R_{\mu\rho\sigma\gamma}R_\nu\,^{\rho\sigma\gamma},R_{\mu\rho\nu\sigma}R^{\rho\sigma}, R_{\mu\nu}R,
 R_{\mu\rho}R_\nu^\rho,  R^2 g_{\mu\nu},
\cr &&R_{\rho\sigma}R^{\rho\sigma}g_{\mu\nu}, R_{\rho\sigma\gamma\eta}R^{\rho\sigma\gamma\eta}
g_{\mu\nu}, \nabla_\mu\nabla_\nu R,\Box R_{\mu\nu},\Box R g_{\mu\nu}\}\,.
\eea
Note that $g_{\mu\nu}$ on the RHS of eq.~\eqref{field redefinition} shall be further scaled by $d_0$ such that the Einstein Hilbert term is invariant under the field redefinition.  (There is no such scaling needed for the Minkowski vacuum.) The consequence is that we have to introduce $\tilde q_i$ to absorb the descendants of the higher-order $c_i$ terms under the field redefinitions.  Under this scheme, the bare $\ell_0$ must also vary under the field redefinition. The explicit rule how Wilson coefficients vary is recorded in Appendix \ref{app: field redefinition} (see eq.~\eqref{eq: field redefinition}), and here we simply quote our findings. The coefficients $q_3$, $c_7$ and $c_8$ are explicitly invariant, and there is one more invariant combination
\be
\tilde q_3'\equiv\tilde{q}_3-4(5c_5+c_6)+\frac{8}{3}(5q_1+q_2)q_3\,.
\ee
We can construct six invariant length parameters
\bea
\label{ellinv D=5}
\ell_{\rm inv}^{(1)}&=&\ell_0-\frac{10(5q_1 +q_2)}{3M^2 \ell_0}-\frac{10(5\tilde{q}_1 +\tilde{q}_2)}{3M^4 \ell_0^3}+\frac{10(4(25c_1+5c_2+c_3+c_4)-3(5q_1 +q_2)^2)}{3M^4\ell_0^3},\cr
\ell_{\rm inv}^{(2)}&=&\frac{q_3}{M^2\ell_0}+\frac{10}{3}\frac{(5q_1+q_2)q_3}{M^4\ell_0^3}
,\qquad \ell_{\rm inv}^{(3)}=\frac{\tilde{q}_3'}{M^4\ell_0^3},\cr
\ell_{\rm inv}^{(4)}&=&\frac{q_3^2}{M^4\ell_0^3}\,,\qquad \ell_{\rm inv}^{(5)}=\frac{c_7}{M^4\ell_0^3}\,,\qquad \ell_{\rm inv}^{(6)}=\frac{c_8}{M^4\ell_0^3}\,.
\eea
Any linear combinations of the above could be the invariant ``AdS radius'' $\ell_{\rm inv}$, as long as the leading term is $\ell_0$. Thus we find that the central charges are indeed invariant, but there is an ambiguity in the expression:
\be
a= \fft{\ell_{\rm inv}^3}{G_N^{(5)}}(\tilde{\#}_1 +\fft{\tilde{\#}_2}{M^2\ell_{\rm inv}^2} + \fft{\tilde{\#}_3}{M^4\ell_{\rm inv}^4})\,,\qquad
c= \fft{\ell_{\rm inv}^3}{G_N^{(5)}}(\tilde{\#}'_1 + \fft{\tilde{\#}'_2}{M^2\ell_{\rm inv}^2} + \fft{\tilde{\#}'_3}{M^4\ell_{\rm inv}^4})\,,
\ee
where $\tilde{\#}_i$ and $\tilde{\#}'_i$ are invariant dimensionless coefficients, depending on the choice of a fiducial length parameter. Part of this ambiguity stems from the choice of the OPE basis for the stress tensor, which we will discuss presently. For now, we simply aim to find a particular $\ell_{\rm inv}$ that validates \eqref{relation 4d1}. The logic is straightforward, we would like to make a field redefinition, bring gravity theory to massless gravity with standard hierarchy $g_{2,i}\sim 1/M^2, g_{3,i}\sim 1/M^4$ (i.e., $\tilde{q}_i=0$). To satisfy the differential relation for massless gravity, $\ell_{\rm inv}$ is fixed to be $\ell_{\rm eff}|_{\rm massless}$. (In this order-by-order approach, we impose the massless condition at each order that gives two $\ell_{\rm eff}$-independent linear relations of the coupling constants. The subtlety issue of $\partial/\partial_{\rm eff}$ of the general Riemann tensor theory does not arise.) This procedure uniquely determines
\bea
\ell_{\rm inv}^{(0)}=(1,\fft{7}{3},\fft{1}{3},-\fft{7}{6},-\fft{26}{3},\fft{11}{2})\cdot\ell_{\rm inv}^{(i)}\,,\label{fix y}
\eea
where the dot $\cdot$ denotes the internal product of two vectors. Adopting this $\ell_{\rm inv}^{(0)}$ for the $(a,c)$ expressions, we find that the invariant dimensionless coefficients are
\begin{eqnarray}
&& \tilde{\#}_1=\tilde{\#}'_1=1\,,\quad \tilde{\#}_2=3\tilde{\#}'_2 =-12 q_3\,,
\cr && \tilde{\#}_3=-3\tilde{\#}'_3 =16 q_3^2-6 \tilde{q}_3' +36c_7-9c_8\,.
\end{eqnarray}
The specific ratios $\tilde{\#}_i/\tilde{\#}_i'$ above lead to the differential relation eq.~\eqref{relation 4d1}.

\begin{figure}[t]
\centering \hspace{0mm}\def\svgwidth{100mm}
\begingroup%
  \makeatletter%
  \providecommand\color[2][]{%
    \errmessage{(Inkscape) Color is used for the text in Inkscape, but the package 'color.sty' is not loaded}%
    \renewcommand\color[2][]{}%
  }%
  \providecommand\transparent[1]{%
    \errmessage{(Inkscape) Transparency is used (non-zero) for the text in Inkscape, but the package 'transparent.sty' is not loaded}%
    \renewcommand\transparent[1]{}%
  }%
  \providecommand\rotatebox[2]{#2}%
  \newcommand*\fsize{\dimexpr\f@size pt\relax}%
  \newcommand*\lineheight[1]{\fontsize{\fsize}{#1\fsize}\selectfont}%
  \ifx\svgwidth\undefined%
    \setlength{\unitlength}{215.53273747bp}%
    \ifx\svgscale\undefined%
      \relax%
    \else%
      \setlength{\unitlength}{\unitlength * \real{\svgscale}}%
    \fi%
  \else%
    \setlength{\unitlength}{\svgwidth}%
  \fi%
  \global\let\svgwidth\undefined%
  \global\let\svgscale\undefined%
  \makeatother%
  \begin{picture}(1,0.21584839)%
    \lineheight{1}%
    \setlength\tabcolsep{0pt}%
    \put(0,0){\includegraphics[width=\unitlength,page=1]{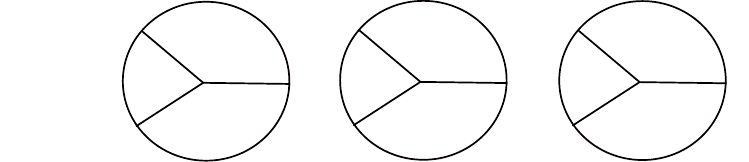}}%
    \put(-0.0018109,0.10161836){\color[rgb]{0,0,0}\makebox(0,0)[lt]{\lineheight{1.25}\smash{\begin{tabular}[t]{l}$\langle TTT\rangle =$\end{tabular}}}}%
    \put(0.25766782,0.11957339){\color[rgb]{0,0,0}\makebox(0,0)[lt]{\lineheight{1.25}\smash{\begin{tabular}[t]{l}$R-2\Lambda$\end{tabular}}}}%
    \put(0.56107411,0.12638672){\color[rgb]{0,0,0}\makebox(0,0)[lt]{\lineheight{1.25}\smash{\begin{tabular}[t]{l}$[R]^{(2)}$\end{tabular}}}}%
    \put(0.84596802,0.12131931){\color[rgb]{0,0,0}\makebox(0,0)[lt]{\lineheight{1.25}\smash{\begin{tabular}[t]{l}$[R]^{(3)}$\end{tabular}}}}%
    \put(0.41125247,0.09124314){\color[rgb]{0,0,0}\makebox(0,0)[lt]{\lineheight{1.25}\smash{\begin{tabular}[t]{l}$+$\end{tabular}}}}%
    \put(0.70100618,0.09391108){\color[rgb]{0,0,0}\makebox(0,0)[lt]{\lineheight{1.25}\smash{\begin{tabular}[t]{l}$+$\end{tabular}}}}%
  \end{picture}%
\endgroup%

\caption{Einstein, quadratic and cubic terms enumerate the stress-tensor three-point structures.}
\label{TTT st}
\end{figure}

It is known in CFTs that central charges can be expressed in terms of $\lambda_{TTT}$, i.e., OPEs of $\langle TTT\rangle$ \cite{Erdmenger:1996yc}. According to conformal symmetry, there are only three parity-even conformal invariant structures in $\langle TTT\rangle$. (In $d=3$, there are only two structures.) The three-point function $\langle TTT\rangle$ is thus determined by the three tensor structures multiplied by the intrinsic OPE $\lambda_{TTT}$. The three structures are enumerated by Einsten, quadratic and cubic gravity vertices, as depicted in Fig \ref{TTT st}. All higher derivative terms either repeat these $\lambda_{TTT}$'s or encode higher point contact structures. For this reason, we may argue that the results we obtain about the OPE coefficients may be universal.

One can think of $\lambda_{TTT}$ as a three-dimensional vector, and there is ambiguity in the choice of the OPE basis, which can explain part of the choices in defining $\ell_{\rm inv}$, since $\lambda_{TTT}$ must be expressible in terms of these invariant lengths. However, there are only three OPEs, but we have five free parameters to define a seemingly valid $\ell_{\rm inv}$. In fact, one can immediately see that the coefficients of $\ell_{\rm inv}^{(2)}$ and $\ell_{\rm inv}^{(3)}$ must be equal in the three-point function from the bulk Witten diagram, because $q_3$ and $\tilde{q}_3$ combine to give $g_{2,3}$. However, the field redefinition appears to allow different coefficients. In other words, there exist $\ell_{\rm inv}$'s that cannot express any OPE by $\ell_{\rm inv}^3$. More specifically, those $\ell_{\rm inv}$'s that can represent OPEs via $\ell_{\rm inv}^3$ have to include the combination
\be
\#(\ell_{\rm inv}^{(2)}+\ell_{\rm inv}^{(3)})-\fft{1}{18}(5+6 \#+18\#^2)\ell_{\rm inv}^{(4)}\,,
\ee
where $\#$ is pure number. To resolve this issue, we note first that at the perturbative order considered, we have $\ell_{\rm inv}^{(4)}=(\ell_{\rm inv}^{(2)})^2/\ell_0$, and therefore, $\ell_{\rm inv}^{(4)}$ should not be considered as an independent length parameter.  Secondly, we may also take the view that the $\tilde q_i$ terms are introduced passively in order to absorb the descendants of the higher order terms from the field redefinition.  We should therefore set the invariant quantity $\tilde q_3'$ to zero to avoid introducing artificial variables. This leads to the precise balancing between three independent length parameters and three OPE coefficients.  In fact we can use the field redefinition from the beginning to remove all terms associated with the Ricci tensor and Ricci scalar, in which case, we have only three nontrivial parameters $(q_3,2c_7+c_8,\ell_{\rm eff})$. (In five bulk dimensions, the cubic Lovelock gravity is trivial for which we have $2c_7+c_8=0$.) Our approach of using $\ell_{\rm inv}$ yields the same result, but makes the procedure more covariant under the field redefinition.

To elaborate this further, we note that the three-point basis proposed in \cite{Erdmenger:1996yc} is often used in the literature, e.g., \cite{Camanho:2013pda,Buchel:2009sk,Myers:2010jv,Bueno:2018xqc,Li:2019auk}, where OPEs are denoted as $\mathcal{A},\mathcal{B},\mathcal{C}$, and in $d=4$ one has \cite{Erdmenger:1996yc}
\bea
&& a=\frac{\pi ^6 }{2880}(13 \mathcal{A}-2 \mathcal{B}-40 \mathcal{C})\equiv \tilde {\mathcal A}\,,
\cr && c=\frac{\pi ^6 }{480} (9 \mathcal{A}-\mathcal{B}-10 \mathcal{C})\equiv \tilde {\mathcal C}\,.\label{acABC}
\eea
OPEs $\mathcal{A},\mathcal{B},\mathcal{C}$ can be explicitly calculated in our example by using ANEC operator \cite{Hofman:2008ar}, which are recorded in Appendix \ref{app:3pt}. These results establish holographically the relations in \eqref{acABC}.

Another natural choice is the orthogonal basis \cite{Caron-Huot:2021kjy,Li:2021snj}, where the orthogonality is defined with respect to three-point pairing \cite{Karateev:2018oml}, because the orthogonal basis in the flat-space limit precisely corresponds to $R$, $[R^{(2)}]$ and $[R^{(3)}]$ respectively. The details of these structures are lengthy and we leave them recorded in Appendix \ref{app:3pt}. In general dimensions $d>3$, we find
\bea
&& \lambda_{TTT}^{(1)}=\frac{-\left(\mathcal{A} \left(d^2+d-2\right)\right)+4 \mathcal{C} (d+1)+2 \mathcal{B}}{d \left(d^2-4\right)}\,,
\cr && \lambda_{TTT}^{(2)}=\frac{2 (d-1) (2 \mathcal{A}+4 \mathcal{C}-\mathcal{B})}{(d-2) d (d+3)}\,,
\cr && \lambda_{TTT}^{(3)}=\frac{\mathcal{A} \left(-d^2+d+4\right)+4 \mathcal{C} (d+1)-2 d \mathcal{B}}{d \left(-d^2+d+2\right)}\,.
\label{OPE rotation}
\eea
In the orthogonal basis, we can immediately observe $\lambda_{TTT}^{(1)}\propto c$. We can thus rotate to a $\ell_{\rm inv}^{\rm orth}$
\be
\ell_{\rm inv}^{\rm orth}=(1,1,1,-\frac{29}{18},-\frac{38}{3},\frac{13}{2})\cdot \ell_{\rm inv}^{(i)}\,,
\ee
for which
\bea
c=\fft{\pi(\ell_{\rm inv}^{\rm orth})^3}{8G_N^{(5)}}=-\fft{\pi^6}{20}\lambda_{TTT}^{(1)}\,.
\eea
The $a$-charge is a bit more complicated, with
\be
\tilde{\#}_1=1\,,\qquad \tilde{\#}_2=-8q_3\,,\qquad
\tilde{\#}_3=\fft{4}{3}(37c_7 -9c_8 -6\tilde{q}_3'+16 q_3^2)\,,
\ee
which satisfies
\be
a=-\fft{\pi^6}{2160}\Big(54 \lambda_{TTT}^{(1)}+35 \lambda_{TTT}^{(2)}+ 15\lambda_{TTT}^{(3)}\Big)\,.
\ee
Regardless the basis, the linear relations among $a$, $c$ and three $\lambda_{TTT}$'s show that we may treat three $\lambda_{TTT}$'s as independent, and $(a,c)$ charges are derived quantities.  The detail coefficients of the relation depend on the basis that we would like to choose.

Hidden relation emerges when we choose a specific invariant length $\ell_{\rm inv}^{(0)}$.
It leads to a linear differential relation between $(a,c)$. We can further establish a new differential relation between $a$-charge and the third linearly independent combination of the OPE coefficients\footnote{We are grateful to Andr\'es Anabalon for pointing out that the $\tilde {\mathcal B}$ in our first version is not independent, but equal to $(c-a)/9$.}
\be
\tilde {\mathcal B}\equiv \fft{\pi^6}{2016}(9 \mathcal{A}-2 \mathcal{B}-2\mathcal{C})
=\Box_{\ell_{\rm inv}^{(0)}} a\,,\label{tildeB}
\ee
where
\be
\nabla_\ell\equiv \ft13 \ell \fft{\partial}{\partial\ell}\,,\qquad \Box_{\ell} \equiv \nabla_{\ell}(\nabla_{\ell}-1)\,.
\ee
(See appendix \ref{app:3pt} for detail derivation.) This relation, together with eq.~\eqref{relation 4d1} have a profound implication: even through stress-tensor OPEs are algebraically independent, they are related to each other via differential operators from the holographic point of view! Furthermore, we have a geometric implication of these differential relations.  The differential operator $\Box_\ell$ is a Laplacian of the four-dimensional hyperbolic space of radius $1/3$, with radial coordinate $\ell$:
\be
ds_4^2=\fft{9(d\ell^2 + dx^2 + dy^2 + dz^2)}{\ell^2}\,,\label{ads4}
\ee
which is Euclidean AdS$_4$ in planar coordinates. In particular, $\nabla_\ell$ is the radial vector of unit length in the hyperbolic space. We expect that there is a systematic way to generate more differential relations among OPEs, and they might be organized under the hidden symmetry associated with the isometry group of the metric \eqref{ads4}, and we leave the construction of this interesting systematics for future study.

\subsection{Further comments: $a$-theorem and functional relations?}

Holographic analysis indicates that there is differential relation between central charges.
It can be established for general Riemann tensor gravity or for perturbative EFT. Perturbatively, the relation is established by differentiating with $\ell_{\rm inv}^{(0)}$ that asymptotes $\ell_0$ in the $M\rightarrow \infty$ limit. However CFTs {\it a priori} do not know the parameter $\ell_0$, it is therefore hard to make sense of the differential relation from the CFT perspective. Here we argue that an analogous differential relation can be established if we extend central charges to radius-dependent functions by deforming away from AdS, exactly following the same approach as addressing the holographic $a$-theorem \cite{Freedman:1999gp}. In this approach, one simply replaces $\ell_{\rm inv}^{(0)}$ by $A'(r)$, where $r\rightarrow\infty$ is the AdS boundary and $A(r)$ flows to $\ell_{\rm inv}^{(0)}$ in pure AdS. We find the following differential relation is valid
\be
\fft{da(r)}{dr}=3c(r)\fft{A''(r)}{A'(r)}\,,\label{relation-function}
\ee
which can be deduced from eq.~\eqref{relation-function} by using the differential chain rule. As was proved in \cite{Osborn:1989td}, at the conformal perturbation level to all loops, one has
\be
\fft{da(\mu)}{d\mu}=G_{ij}\beta^i\beta^j\,,
\ee
where $G_{ij}>0$ and $\beta^i$s are $\beta$-functions.  Here $\mu$ is the energy scale that flows to UV and corresponds to $r$ in the AdS bulk. This is exactly the $a$-theorem, and we conclude
\be
G_{ij}\beta^i\beta^j=3c(\mu)\fft{A''(\mu)}{A'(\mu)}\,,
\ee
which is consistent because of $A''(r)|_{\rm AdS}=0$. Since $c(r)>0$ for any unitary theory, we thus find holographic $a$-theorem is validated provided $A''(r)/A'(r)>0$. It is interesting to study the relation between $A''(r)/A'(r)>0$ and the null energy condition in a generic AdS EFT in the future. More nontrivially, eq.~\eqref{tildeB} then implies a novel differential relation among the OPE $\tilde{\mathcal B}$-function and the $(a,c)$ functions, which are themselves OPE functions:
\be
c'(\mu) - a'(\mu) = \fft{a'(\mu)}{c(\mu)}\tilde{\mathcal B}(\mu)\,.
\ee

\section{Conclusion}

Two related topics were addressed in this paper. One is that we proved the differential relation \eqref{relation 4d} of \cite{Li:2018drw} and its higher-dimensional generalization for massless gravities that are constructed from the most general Riemann tensor invariants.
The second is to approach this relation perturbatively, order by order.  This allowed us to find more differential relations among the OPE coefficients of the three-point function $\langle TTT\rangle$ of the stress tensor.

We considered AdS gravity extended with general Riemann tensor invariants up to and including the cubic order.  We showed that the $(a,c)$ central charges and three OPE coefficients $({\cal A}, {\cal B}, {\cal C})$ were all invariant under field redefinitions of the metric.  We reconfirmed the known fact that both central charges could be expressed as certain linear combinations of the OPE coefficients.  By recombining these OPE coefficients to $(\tilde {\cal A}, \tilde {\cal B}, \tilde {\cal C})$, we find that two of them can be expressed as differentials of the third:
\be
\tilde {\mathcal{C}}=\nabla_{\ell_{\rm inv}^{(0)}} \tilde {\mathcal{A}}\,,\qquad
\tilde {\mathcal{B}} = \Box_{\ell_{\rm inv}^{(0)}} \tilde {\mathcal{A}}\,.\label{hidden}
\ee
where $\nabla_{\ell}$ and $\Box_{\ell}$ are the unit radial vector and the Laplacian of the four-dimensional hyperbolic space \eqref{ads4} respectively. The central charge differential relation \eqref{relation 4d1} is the consequence of the first relation above. However, there is a difference: The relations in \eqref{hidden} are perturbative results since OPE coefficients can only defined perturbatively. The central-charge relation \eqref{relation 4d1}, on the other hand, can be promoted to \eqref{relation 4d} for general massless gravities whose higher-order couplings are not necessarily small. The hidden relations tantalizingly suggest that there may exist a deep organizing principle for the correlation functions of CFTs, associated with the isometry group of the metric \eqref{ads4}, which can be viewed as Euclidean AdS$_4$ in planar coordinates.

However, the understanding of the differential relations from the CFT perspective remains illusive, since $\ell_{\rm inv}^{(0)}$ does not have an immediate translation to any quantity in CFTs. We argue that we can use the analogous approach in the holographic $a$-theorem and obtain the corresponding differential relations of $(a,c)$ functions but also those of the OPE coefficients. However, the subject requires further investigation.

\section{Acknowledgement}

We would like to thank Zhan-Feng Mai for useful conversation and initial collaboration on proof of relation of central charges, and Simon Caron-Huot, Alexander Maloney for discussions. The work is supported in part by the National Natural Science Foundation of China (NSFC) grants No.~11875200 and No.~11935009. Y.-Z.L.~is also supported in part by the Fonds de Recherche du Qu\'ebec - Nature et Technologies and by the Simons Collaboration on the Nonperturbative Bootstrap. H.L.~and L.M.~benefit also from NSFC grants No.~11947301 and No.~12047502.

\section*{Appendix}
\appendix

This Appendix includes: generalization of the relation of central charges to general dimensions (sec.\,A), details of field redefinition of Wilson coefficients up to cubic order in general dimensions (Sec.\,B), three-point structures of $\langle TTT\rangle$ in general dimensions and the calculation of corresponding OPEs from ANEC operator in $d=4$ (Sec.\,C).

\section{Relation of central charges in general dimensions}
\label{app: relation in general dim}
\subsection{Holographic central charges}

The differential relation of the central charges can be generalized to general dimensions $d\geq 3$. However, there is a subtlety since in higher dimensions, there are more central charges and they can be very complicated. Nevertheless, the central charge $C_T$ and $a$ can be universally defined.  (The identification of $C_T$ with a suitable $c$ was only done holographically in \cite{Lu:2019urr}). It is also important to note even though there is no conformal anomaly in odd $d$, $C_T$ is always well-defined
\bea
\label{relation between CT and keff}
C_T=\frac{\pi ^{-\frac{d}{2}-1} \Gamma (d+2)}{8 (d-1) \Gamma \left(\frac{d}{2}\right)}
\frac{\ell_{\mathrm{eff}}^{d-1}}{G_{N\mathrm{eff}}^{(d+1)}}.
\eea
The substitute of $a$-charge in odd $d$ is the entanglement entropy (EE) for spherical entangling surface $S^{d-2}$ \cite{Casini:2011kv}. It is shown that EE for spherical entanglement surface can be mapped to thermal entropy of $R\times H^{d-1}$ via an appropriate conformal map.  In holography, EE for half of $S^{d-1}$ of $R\times S^{d-1}$ boundary of global AdS boils down to black hole entropy of hyperbolic topological black hole in $d+1$-dimension \cite{Casini:2011kv}
\be
ds^2_{d+1}=-fdt^2+\frac{dr^2}{f}+r^2d\Omega^2_{d-1,k=-1},\ \ \ f=\frac{r^2}{\ell_{\mathrm{eff}}^2}-1\,,
\ee
which is locally AdS.  According to the Wald formalism \cite{Wald:1993nt}, we have
\be
S=-\fft{\omega_{k,d-1}\ell_{\rm eff}^{d-1}}{8G_N^{(d+1)}}\Big(P_{abcd}\epsilon^{ab}\epsilon^{cd}   \Big)_{r=\ell_{\rm eff}}\,,
\ee
where $\bar{P}_{abcd}\epsilon^{ab}\epsilon^{cd}=-4\xi_0$ and $\omega_{k,d-1}$ denotes the entangling surface and it is divergent in even $d$, . It is instructive to simply divide by $\omega_{k,d-1}$ and define the density
\be
\label{a charge entanglement}
a^\ast=\frac{\Omega_{d-1}}{4\pi\omega_{k,d-1}}S=\frac{\pi ^{\frac{d}{2}-1} \xi _0\ell_{\rm eff}^{d-1}}{4 \Gamma \left(\frac{d}{2}\right)G_N^{(d+1)}}\,,
\ee
where $\Omega_{d-1}$ is the volume of $d-1$-sphere. It is clear from the Wald formalism that $a^*$ does not explicitly depend on the bare cosmological constant, which we solve for in terms of $\ell_{\rm eff}$ and other coupling constants in the theory.

In even $d$, this reproduces a known fact that EE for spherical entangling surface is proportional to $a$-charge. To show this, we apply the reduced the FG trick in general even dimensions by restricting to $S^{d-1}$ boundary topology.
\bea
\label{reduce FG}
ds^2=\frac{\ell_{\mathrm{eff}}^2}{4\rho^2}d\rho^2+\frac{f(\rho)}{\rho}d\Omega_{d}^2\,,
\eea
which kills all the Weyl invariants. We expand the action
\bea
S=\frac{1}{16\pi G_N^{(d+1)}}\int d^{d+1}x\sqrt{-\hat{g}_{d+1}}\mathcal{L}
\eea
around the AdS boundary $\rho\rightarrow0$
\bea
\mathcal{L}&&=\mathcal{L}_{\mathrm{vac}}+\mathcal{L}_1\big|_{\rho\rightarrow0}\rho+\mathcal{L}_2\big|_{\rho\rightarrow0}\rho^2+\cdots+\mathcal{L}_{\frac{d}{2}}\big|_{\rho\rightarrow0}\rho^\frac{d}{2},\\
\sqrt{-\hat{g}_{d+1}}&&=\frac{\ell_{\mathrm{eff}}}{2}\rho^{-\frac{d+2}{2}}\left(a_0+a_1\rho+a_2\rho^2+\cdots+a_{\frac{d}{2}}\rho^\frac{d}{2}\right)\sim\frac{\ell_{\mathrm{eff}}}{2}\rho^{-\frac{d+2}{2}}f^{\frac{d}{2}}
\eea
and then collect all the terms contributing to $\rho^{-1}$. The ansatz for $f$ is
\bea
f=&&f_0+f_1\rho+f_2\rho^2+f_3\rho^3+\cdots+f_{\frac{d}{2}}\rho^\frac{d}{2}.
\eea
Performing variation with respect to the coefficients $f_i$ and then applying the variation principle, we find
\be
f_1=-\frac{\ell_{\mathrm{eff}}^2}{2},\qquad f_2=\frac{\ell_{\rm eff}^4}{16f_0},\qquad  f_i=0,\qquad 3\leq i\leq \frac{d-2}{2}.
\ee
Thus we have
\bea
\mathcal{L}&=&\mathcal{L}_{\mathrm{vac}}+\mathcal{L}_\frac{d}{2}\rho^{\frac{d}{2}},\qquad \mathcal{L}_{\mathrm{vac}}=-\frac{4d}{\ell_{\mathrm{eff}}^2}\xi_0,\qquad \mathcal{L}_\frac{d}{2}=\frac{2d^2}{\ell_{\mathrm{eff}}^2}\frac{f_{\frac{d}{2}}}{f_0}\xi_0,\\
a_{\frac{d}{2}}&=&-\frac{\ell_{\mathrm{eff}}^{d}}{d}-\frac{(-4)^{\frac{d}{2}}}{2}\frac{\Gamma(\fft{d+2}{2})^2}{\Gamma(d+1)}
f_{\frac{d}{2}}f_0^{\frac{d-2}{2}},\qquad a_0=-\frac{(-4)^{\frac{d}{2}}}{d}\frac{\Gamma(\fft{d+2}{2})^2}{\Gamma(d+1)}f_0^{\frac{d}{2}}.
\eea
We find precisely
\bea
a=a^\ast\,,\quad \text{for even}\,\, d\,.
\eea
It is of interest to note that from this action procedure to calculate the $a$-charge, we need to know explicitly the on-shell condition $\mathcal{L}_{\mathrm{vac}}= -\frac{4d}{\ell_{\mathrm{eff}}^2} \xi_0$, while the equation \eqref{a charge entanglement} does not explicitly involve this relation.

\subsection{Complete the proof of the relation}

The proof is analogous to the $D=5$ example presented in the main text. It follows from  (\ref{EOM AdS dh}) and (\ref{Gneff}) that we have
\be
\frac{1}{G_{N\mathrm{eff}}^{(d+1)}}=\frac{\ell_{\mathrm{eff}}^3}{2d(d-1)G_N^{(d+1)}}\left(\mathfrak{h}'
(\ell_{\mathrm{eff}})+\frac{8d^2}{\ell_{\mathrm{eff}}^5}\xi\right),
\ee
where $\xi\equiv (d-3)\xi_1-2(d+1)\xi_2-2\xi_3$. We therefore have
\be
C_T=\frac{\Omega_{d-1}\Gamma(d+2)}{16\pi^{d+1}(d-1)}\frac{\ell_{\mathrm{eff}}^{d-1}}
{G_{N\mathrm{eff}}^{(d+1)}}
=\frac{\Omega_{d-1}\Gamma(d+2)\ell_{\mathrm{eff}}^{d+2}}{32\pi^{d+1}d(d-1)^2G_N^{(d+1)}}\left(\mathfrak{h}'
(\ell_{\mathrm{eff}})+\frac{8d^2}{\ell_{\mathrm{eff}}^5}\xi\right).
\ee
Taking a derivative of eq.~\eqref{EOM AdS h} gives:
\bea
\label{ac relation}
C_T&=&\frac{\Omega_{d-1}\Gamma(d+2)}{32\pi^{d+1}d(d-1)^2G_N^{(d+1)}}
\Big[\ell_{\mathrm{eff}}^{d+2}\bar{\mathcal{L}}'(\ell_{\mathrm{eff}})-
8\xi_0d\ell_{\mathrm{eff}}^{d-1}+4d\ell_{\mathrm{eff}}^{d}\frac{\partial \xi_0}{\partial\ell_{\mathrm{eff}}}+8d^2\ell_{\mathrm{eff}}^{d-3}\xi  \Big]\cr
&=&\frac{\Gamma(d+2)}{\pi^d(d-1)^2}\Bigg\{\ell_{\mathrm{eff}}
\frac{\partial}{\partial\ell_{\mathrm{eff}}}\left[\frac{\Omega_{d-1}\xi_0
\ell_{\mathrm{eff}}^{d-1}}{8\pi G_N^{(d+1)}}\right]
+\frac{\Omega_{d-1}d}{4\pi G_N^{(d+1)}}\ell_{\mathrm{eff}}^{d-3}\xi\Bigg\}\cr
&=&\frac{\Gamma(d+2)}{\pi^d(d-1)^2}\ell_{\mathrm{eff}}\frac{\partial a^\ast}{\partial\ell_{\mathrm{eff}}}+
\frac{\Omega_{d-1}\Gamma(d+2)d}{4\pi^{d+1}(d-1)^2G_N^{(d+1)}}\xi\,.
\eea
After imposing the massless conditions that imply $\xi=0$, we obtain the universal central charge relation
\bea
C_T=\frac{\Gamma(d+2)}{\pi^d(d-1)^2}\ell_{\mathrm{eff}}\frac{\partial a^\ast}{\partial\ell_{\mathrm{eff}}}.
\eea
Again we should emphasize that the derivative with respect to $\ell_{\rm eff}$ should be done before imposing the massless conditions such that all the coupling constants are independent of $\ell_{\rm eff}$.  We shall illustrate this with two explicit examples in the next subsection.

This differential relation was first proposed in \cite{Li:2018drw}, and was independently noticed in \cite{Bueno:2018yzo}, where $a^\ast$ was replaced by a quantity from the free energy of sphere. As was explained in the main text, from the EFT approach of AdS gravity, we only have to replace $\ell_{\rm eff}$ by a suitable $\ell_{\rm inv}^{(0)}$ and the differential relation holds. We record $\ell_{\rm inv}^{(0)}$ for the cubic example in appendix B.

\subsection{Two explicit examples}

Here, we address the subtleties of taking a derivative with respect to $\ell_{\rm eff}$ using two explicit examples. In this paper we consider a general class of theories of the type
\be
{\cal L} = R -2 \Lambda_0 + \tilde {\cal L}(R_{\mu\nu\rho\sigma}, g^{\mu\nu}, \alpha)\,,
\ee
where $\alpha$ denotes all the dimensionful coupling constants of the Riemann tensor invariants.  We assume that the theory admits an AdS vacuum of radius $\ell_{\rm eff}$ and the on-shell condition, i.e.~the first equation in \eqref{EOM AdS L}.  This enables us to express $\Lambda_0$ as a function of $\ell_{\rm eff}$ and $\alpha$ which is independent of $\ell_{\rm eff}$. Our derivative with respect to $\ell_{\rm eff}$ is thus taken such that $\partial \alpha/\partial \ell_{\rm eff}=0$.  (In fact, the derivatives of various quantities with respect to $\ell_{\rm eff}$ do not involve $\Lambda_0$ explicitly either.) The massless conditions, on the other hand, relate coupling constants that could involve $\ell_{\rm eff}$ and therefore the conditions should be imposed after take the derivatives.

\subsubsection{A polynomial example}

We first consider the Einstein-quadratic-cubic-AdS theory in five dimensions $(d=4)$. The theory was studied in detail in section \ref{section EFT} as an effective theory.  Here we treat it as a classical theory where the coupling constants are not necessarily small.
The three quadratic and eight cubic Riemann tensor invariants are summarized in \eqref{quadratic cubic theory}. We use $(q_1,q_2,q_3)$, $(c_1,c_2,\ldots, c_8)$ to denote the quadratic and cubic coupling constants respectively.  We have
\bea
\Lambda_0&=&-\frac{6}{\ell _{\text{eff}}^2}+\frac{4 \left(10 q_1+2 q_2+q_3\right)}{\ell _{\text{eff}}^4}\nn\\
&&+\frac{2 \left(400 c_1+80 c_2+16 c_3+16 c_4+40 c_5+8 c_6+4 c_7+3 c_8\right)}{\ell _{\text{eff}}^6},
\eea
and $(a,c)$ central charges
\bea
a&=&\frac{\pi}{8G_N^{(5)}}\Big(\ell _{\text{eff}}^3-4 \left(10 q_1+2 q_2+q_3\right) \ell _{\text{eff}}\cr
&&+\frac{3 \left(400 c_1+80 c_2+16 c_3+16 c_4+40 c_5+8 c_6+4 c_7+3 c_8\right)}{\ell _{\text{eff}}}\Big),\nn\\
c &=&\frac{\pi}{8G_N^{(5)}}\Big(\ell _{\text{eff}}^3-4 \left(10 q_1+2 q_2-q_3\right) \ell _{\text{eff}}\cr
&&+\frac{1200 c_1+240 c_2+48 c_3+48 c_4-40 c_5-8 c_6-36 c_7+21 c_8}{\ell _{\text{eff}}}\Big).
\eea
Taking a derivative of $a$ with respect to $\ell_{\rm eff}$, while treating $q_i$ and $c_i$ couplings as independent, we have
\bea
c&=&\fft13 \ell_{\rm eff} \fft{\partial a}{\partial \ell_{\rm eff}} + \fft{\pi}{3G_N^{(5)} \ell_{\rm eff}}\, \xi\,,\nn\\
\xi &=& 3(200 c_1 + 40 c_2 + 8 c_3 + 8 c_4 - 4 c_7 + 3 c_8) - 2 (5q_1 + q_2 - q_3)\ell_{\rm eff}^2\,.
\eea
From the massless conditions
\bea
q_3&=&-q_1-\frac{q_2}{2}+\frac{3}{2 \ell_{\mathrm{eff}} ^2}(40 c_1+12 c_2+4 c_3+3 c_4+16 c_5+4 c_6+4 c_7),\cr
c_8&=&\frac{\ell_{\mathrm{eff}} ^2 }{3}\left(4 q_1+q_2\right) -\frac{1}{3}(240 c_1+52 c_2+12 c_3+11 c_4+16 c_5+4 c_6),
\eea
we can easily check that $\xi=0$, giving rise to \eqref{relation 4d}. In the past literature \cite{Li:2017txk,Li:2018drw} the massless conditions are impose at each order independently, such that one has two linear relations for $q_i$'s, as well as for $c_i$'s, which then do not involve $\ell_{\rm eff}$.  One can therefore imposing the massless conditions before taking the $\ell_{\rm eff}$ derivatives in these cases. In the next, we present a nontrivial example where the massless conditions necessarily involve $\ell_{\rm eff}$.

\subsubsection{A fractional example}

Here we consider a more complicated fractional theory in five dimensions, with the Lagrangian

\be
\mathcal{L}=R-2\Lambda_0+\frac{\alpha}{1+\beta_1R^2+\beta_2R_{\mu\nu}R^{\mu\nu}+
\beta_3R_{\mu\nu\rho\sigma}R^{\mu\nu\rho\sigma}}\,.
\ee
The on-shell condition for the bare cosmological constant is
\bea
\Lambda_0=-\frac{6}{\ell_{\mathrm{eff}} ^2}+\frac{\alpha  \ell_{\mathrm{eff}} ^4 \left(720 \beta _1+144 \beta _2+72 \beta _3+\ell_{\mathrm{eff}} ^4\right)}{2 \left(400 \beta _1+80 \beta _2+40 \beta _3+\ell_{\mathrm{eff}} ^4\right){}^2}.
\eea
Using the techniques outlined in the main text, we can easily obtain the $(a,c)$ central charges:
\bea
a&=&\frac{\pi}{8G_N^{(5)}}\left(\ell_{\mathrm{eff}} ^3+\frac{4 \alpha  \left(10 \beta _1+2 \beta _2+\beta _3\right) \ell_{\mathrm{eff}} ^9}{\left(\ell_{\mathrm{eff}} ^4+40 \left(10 \beta _1+2 \beta _2+\beta _3\right)\right)^2}\right),\cr
c&=&\frac{\pi}{8G_N^{(5)}}\left(\ell_{\mathrm{eff}} ^3+\frac{4 \alpha  \left(10 \beta _1+2 \beta _2-\beta _3\right) \ell_{\mathrm{eff}} ^9}{\left(\ell_{\mathrm{eff}} ^4+40 \left(10 \beta _1+2 \beta _2+\beta _3\right)\right)^2}\right).
\eea
Again these can be obtained without explicitly solving for $\Lambda_0$.  We therefore have
\bea
c&=&\frac{1}{3}\ell_{\mathrm{eff}}\frac{\partial a}{\partial\ell_{\mathrm{eff}}} +\frac{2\pi \alpha  \ell_{\mathrm{eff}} ^9}{3G_N^{(5)}\left(\ell_{\mathrm{eff}} ^4+40 \left(10 \beta _1+2 \beta _2+\beta _3\right)\right)^3}\, \xi\,,\nn\\
\xi &=&\left(\beta_3-5 \beta _1-\beta _2\right) \ell_{\mathrm{eff}} ^4+120 \left(50 \beta _1^2+5 \left(4 \beta _2+3 \beta _3\right) \beta _1+2 \beta _2^2+\beta _3^2+3 \beta _2 \beta _3\right).
\eea
In the above, the derivative with respect to $\ell_{\rm eff}$ was taken by treating $(\alpha, \beta_i)$ parameters as being independent of $\ell_{\rm eff}$. The massless conditions are
\be
\beta_2=-4\beta_3,\quad \beta_1=\frac{\ell_{\mathrm{eff}} ^4+1560 \beta _3+\sqrt{\ell_{\mathrm{eff}} ^8-1680 \beta _3 \ell_{\mathrm{eff}} ^4+14400 \beta _3^2}}{2400}.
\ee
We find that $\xi$ vanishes identically under these conditions, giving rise to \eqref{relation 4d1}.  In this example, we see that the massless condition will necessarily make the couplings depending on $\ell_{\rm eff}$ and therefore, the derivative must be taken before the massless condition.

\section{Field redefinitions and invariant central charges}
\label{app: field redefinition}
Here, we consider the perturbative approach to AdS gravity in general dimensions. We consider Einstein gravity with a negative bare cosmological constant, extended with Riemann tensor invariants \eqref{genlag}, up to and including six derivatives total.

\subsection{Effective AdS}

First, we present the general holographic central charges of effective AdS of radius $\ell_{\mathrm{eff}}$. The equations of motion relate the bare cosmological constant $\Lambda_0$ and $\ell_{\mathrm{eff}}$
\bea
\Lambda_0&=&-\frac{d(d-1) }{2 \ell_{\mathrm{eff}} ^2}+\frac{ d(d-3)}{2 \ell_{\mathrm{eff}} ^4}\Big( d (d+1)g_{2,1}+dg_{2,2}+2g_{2,3}\Big)-\frac{d(d-5) }{2 \ell_{\mathrm{eff}} ^6}\Big(d^2(d+1)^2g_{3,1} \cr
&&+d^2(d+1)g_{3,2}+d^2g_{3,3}+d^2g_{3,4}+2d(d+1)g_{3,5}+2dg_{3,6}+4g_{3,7}+(d-1)g_{3,8}\Big).
\eea
$a$-charge and $C_T$ charge are explicitly given as follows \cite{Li:2017txk}
\bea
a&=&\frac{\Omega_{d-1}}{16\pi G_N^{(d+1)}}\Big(\ell_{\mathrm{eff}}^{d-1}-2\Big(d(d+1)g_{2,1}+dg_{2,2}+2g_{2,3}\Big)\ell_{\mathrm{eff}}^{d-3}+3
\big(d^2(d+1)^2g_{3,1}\cr
&&+d^2(d+1)g_{3,2}+d^2g_{3,3}+d^2g_{3,4}+2d(d+1)g_{3,5}+2dg_{3,6}+4g_{3,7}+(d-1)g_{3,8}\big)\ell_{\mathrm{eff}}^{d-5}\Big),\cr && \\
C_T&=&\frac{\Omega_{d-1}\Gamma(d+2)}{16\pi^{d+1}(d-1)G_N^{(d+1)}}\Big(\ell_{\mathrm{eff}}^{d-1}
-2\big(d(d+1)g_{2,1}+dg_{2,2}-2(d-3)g_{2,3}\big)
\ell_{\mathrm{eff}}^{d-3}\cr
&&+\big(3 d^2 (d+1)^2 g_{3,1}+3 d^2 (d+1) g_{3,2}+3d^2g_{3,3}+3d^2 g_{3,4}-2d(d+1) (2d -7)g_{3,5}\cr
&&-2d (2d -7)g_{3,6} -12 (2d -5)g_{3,7} +3(3d-5)g_{3,8}
\big)\ell_{\mathrm{eff}}^{d-5}\Big)\,.
\eea
In $d=4$, we have $C_T=40/\pi^4 c$.

\subsection{Field redefinitions}

The general field redefinition that is relevant to the cubic-order of curvature tensor is given by \eqref{field redefinition} with
\be
d_0 =1+\frac{\alpha_0}{M^2\ell_0^2}+\frac{\beta_0}{M^4\ell_0^4}\,,\qquad d_{1i}=\fft{\alpha_i}{M^2}+\fft{\tilde{\alpha}_i}{M^4\ell_0^2}\,,\qquad
d_{2i}=\fft{\beta_i}{M^4}\,.
\ee
Note that the scaling coefficient $d_0$ of the metric is to ensure that the Einstein Hilbert term is invariant under the field redefinition.  This requires that
\be
\alpha_0=-d\left(\alpha_1+(d+1)\alpha_2\right),\ \ \ \beta_0=-d(\tilde{\alpha}_1+(d+1)\tilde{\alpha}_2)-\frac{1}{4}d^2(d+1)(\alpha_1+(d+1)\alpha_2)^2,
\ee
The bare cosmological constant is not invariant under the field redefinition; it is shifted by
\be
\Lambda_0\rightarrow \tilde{\Lambda}=\Lambda_0+\frac{\Lambda_{\mathrm{q}}}{M^2\ell_0^2}+
\frac{\Lambda_{\mathrm{c}}}{M^4\ell_0^4}\,,
\ee
with
\bea
&&\Lambda_{\mathrm{q}}=-\frac{d^2(d^2-1)(\alpha_1+(d+1)\alpha_2)}{4\ell_0^2},\cr
&&\Lambda_{\mathrm{c}}=-\frac{d^2(d^2-1)(\tilde{\alpha}_1+(d+1)
\tilde{\alpha}_2)}{4\ell_0^{2}}-
\frac{d^3(d^2-1)(d+2)(\alpha_1+(d+1)\alpha_2)^2}{8\ell_0^2}\,.
\eea
Thus we have
\bea
\ell_0&\rightarrow&  \ell_0+\frac{d(d+1)}{4M^2\ell_0}(\alpha_1+(d+1)\alpha_2)-\frac{d^2(d-3)(d+1)}
{32M^4\ell_0^3}(\alpha_1+(d+1)\alpha_2)^2\cr
&&+\frac{d(d+1)}{4M^4\ell_0^3}(\tilde{\alpha}_1+(d+1)\tilde{\alpha}_2)\,.
\eea
Note that the coefficients $g_{n,i}$ are given by \eqref{gnicoeff}. The rule of the field redefinition for each coupling coefficients $g_{n,i}$ is explicitly presented below
\bea
\label{coefficients shift}
q_1&\rightarrow& q_1+\frac{1}{2}\alpha_1+\frac{d-1}{2}\alpha_2,\qquad
q_2\rightarrow  q_2-\alpha_1,\cr
\tilde{q}_1&\rightarrow&  \tilde{q}_1+\frac{1}{2}\tilde{\alpha}_1+\frac{d-1}{2}\tilde{\alpha}_2+\frac{d(d-1)}{2}\beta_3+\frac{d(d-1)(d+1)}{2}
\beta_5-\frac{d(d-3)}{2}\alpha_1q_1\cr
&&-\frac{d(d-5)}{8}\alpha_1^2 -\frac{d(d-3)(d+1)}{2}\alpha_2q_1-\frac{d(d^2-4d-1)}{4}\alpha_1\alpha_2\cr
&& -\frac{d(d-1)(d-5)(d+1)}{8}\alpha_2^2\,,
\cr
\tilde{q}_2&\rightarrow& \tilde{q}_2-\tilde{\alpha}_1+\frac{d(d-1)}{2}\beta_2+\frac{d(d-1)}{2}\beta_4+\frac{d(d-1)(d+1)}{2}
\beta_6\cr
&&-\frac{d(d-3)}{2}\alpha_1q_2+\frac{d(d-5)}{4}\alpha_1^2 -\frac{d(d-3)(d+1)}{2}\alpha_2q_2+\frac{d(d-3)(d+1)}{2}\alpha_1\alpha_2
\cr
\tilde{q}_3&\rightarrow& \tilde{q}_3+\frac{d(d-1)}{2}\beta_1+\frac{d(d-1)(d+1)}{2}\beta_7-\frac{d(d-3)}{2}(\alpha_1+(d+1)\alpha_2)q_3,
\cr
c_1&\rightarrow& c_1+\frac{1}{2}\beta_3+\frac{d-1}{2}\beta_5+\frac{1}{2}\alpha_1q_1+\frac{1}{8}\alpha_1^2+\frac{d-3}{2}\alpha_2q_1
+\frac{d-3}{4}\alpha_1\alpha_2+\frac{(d-1)(d-3)}{8}\alpha_2^2,
\cr
c_2&\rightarrow&  c_2+\frac{1}{2}\beta_2-\beta_3+\frac{1}{2}\beta_4+\frac{d-1}{2}\beta_6-\frac{1}{2}\alpha_1(4q_1-q_2)-\frac{3}{4}\alpha_1^2
+\frac{d-3}{4}\alpha_2q_2-\frac{d-3}{2}\alpha_1\alpha_2,
\cr
c_3&\rightarrow&  c_3-\beta_4+4\alpha_1q_3+\alpha_1^2,\qquad
c_4\rightarrow  c_4-\beta_2-2\alpha_1(q_2+2q_3),
\cr
c_5&\rightarrow& c_5+\frac{1}{2}\beta_1+\frac{d-1}{2}\beta_7+\frac{1}{2}(\alpha_1+(d-3)\alpha_2)q_3,\qquad
c_6\rightarrow c_6-\beta_1-2\alpha_1q_3,
\cr
c_9&\rightarrow& c_9-\beta_9-\alpha_1(q_2+4q_3),
\cr
c_{10}&\rightarrow& c_{10}+\frac{1}{2}\beta_9+\frac{d-1}{2}\beta_{10}-(\alpha_1+2d\alpha_2)q_1-
\frac{d+1}{2}\alpha_2q_2+\alpha_1q_3-2\alpha_2q_3.\label{eq: field redefinition}
\eea
It is not hard to use this rule to bring any cubic gravity to, e.g., quasi-topological gravity and Einsteinian cubic gravity, analogous to flat-space \cite{Bueno:2019ltp}.

Invariant quantities include $q_3, c_7, c_8$, together with
\be
\tilde{q}_3'=\tilde{q}_3-d((d+1)c_5+c_6)+
\frac{2d(d-3)}{d-1}((d+1)q_1+q_2)q_3\,.
\ee
The $d=4$ case was present in the main text.
\subsection{Relation of central charges}

In the previous subsection, we obtain the transformation rules for each coupling coefficients, including the bare cosmological constant, under the general field redefinition. We are interested in showing that the physical quantities such as central charges are indeed invariant under the field redefinitions. It is clear that the coupling coefficients $q_3$, $c_7$, $c_8$ are manifestly invariant.  For general extended gravity up to and including the cubic orders, we find a total of six field-redefinition invariant lengths $\ell_{\rm inv}^{(i)}$
\bea
\label{ellinv0-5}
\ell^{(1)}_{\mathrm{inv}}&=&\ell_0-\frac{ d(d+1)}{2 (d-1)}\frac{((d+1)q_1 +q_2)}{M^2 \ell_0}-\frac{ d(d+1)}{2 (d-1)}\frac{((d+1)\tilde{q}_1 +\tilde{q}_2)}{ M^4\ell_0^3}\cr
&&+\frac{d^2(d+1) }{2 (d-1)}\frac{(d+1)^2c_1+(d+1)c_2+c_3+c_4-\frac{(5 d-11) }{4 (d-1)}((d+1)q_1 +q_2)^2
}{M^4\ell_0^3},\cr
\ell^{(2)}_{\mathrm{inv}}&=&\frac{q_3}{M^2\ell_0}+\frac{d(d+1)}{2(d-1)}
\frac{((d+1)q_1+q_2)q_3}{M^4\ell_0^3},\qquad \ell^{(3)}_{\mathrm{inv}}=\frac{\tilde{q}_3'}{M^4\ell_0^3},\cr
\ell^{(4)}_{\mathrm{inv}}&=&\frac{q_3^2}{M^4\ell_0^3}\,,\qquad \ell^{(5)}_{\mathrm{inv}}=\frac{c_7}{M^4\ell_0^3}\,,\qquad \ell^{(6)}_{\mathrm{inv}}=\frac{c_8}{M^4\ell_0^3}\,.
\eea
From the perturbative point of view, the length scale $\ell_{\mathrm{inv}}$ can be chosen to be $\ell^{(1)}_{\mathrm{inv}}$ plus a linear combination of $\ell^{(i)}_{\mathrm{inv}}$, ($i=2,3,4,5,6$,) with any numerical coefficients. Regardless the choice of $\ell_{\mathrm{inv}}$, we find that the central charges can always be expressed as
\bea
&& a= \frac{\Omega_{d-1}}{16\pi G_N^{(d+1)}}\left(\ell_{\rm inv}^{d-1} \tilde{\#}_1 + \ell_{\rm inv}^{d-3} \frac{\tilde{\#}_2}{M^2} + \ell_{\rm inv}^{d-5} \frac{\tilde{\#}_3}{M^4}\right)\,,
\cr && C_T= \frac{\Omega_{d-1}\Gamma(d+2)}{16\pi^{d+1}(d-1)G_N^{(d+1)}}\left(\ell_{\rm inv}^{d-1} \tilde{\#}'_1 + \ell_{\rm inv}^{d-3} \frac{\tilde{\#}'_2}{M^2} + \ell_{\rm inv}^{d-5} \frac{\tilde{\#}'_3}{M^4}\right)\,,
\eea
where the coefficients $(\tilde{\#}_i,\tilde{\#}'_i)$ are invariant under the field redefinition, thereby proving that the central charges are invariant.

There exists a particular choice of $\ell_{\mathrm{inv}}$, which we name $\ell_{\rm inv}^{(0)}$. For this choice, $\ell_{\rm inv}^{(0)}$ reduces to $\ell_{\rm eff}$ when the field redefinition brings us to massless gravity. It is given by
\bea
\ell_{\rm inv}^{(0)}&&=(y_1,y_2,y_3,y_4,y_5,y_6)\cdot\ell_{\rm inv}^{(i)}\\
y_1&=&1\,\qquad y_2=\frac{(d-3)(2d-1)}{d-1}\,,\qquad y_3=\frac{d^2-4d+1}{d-1},\cr
y_4&=&-\frac{(d-3)^2(2d-5)(2d-1)}{2(d-1)^2}\,,\quad y_5=-\frac{2(3d^2-10d+5)}{d-1}\,,\quad y_6=\frac{1}{2}(4d-5)\,.
\eea
In terms of $\ell_{\rm inv}^{(0)}$, we find
\begin{eqnarray}
&& \tilde{\#}_1=\tilde{\#}'_1=1\,,\quad \tilde{\#}_2=\frac{d-1}{d-3}\tilde{\#}'_2 =-2(d-1)(d-2) q_3\,,
\cr && \tilde{\#}_3=\frac{d-1}{d-5}\tilde{\#}'_3 =\frac{d-2}{2}\Big(12(d-1)c_7-3(d-1)c_8-2(d-1) \tilde{q}_3'+4 d(d-3)^2q_3^2\Big)\,,
\end{eqnarray}
which manifestly satisfies the relation
\bea
C_T=\frac{\Gamma(d+2)}{\pi^d(d-1)^2}\ell^{(0)}_{\mathrm{inv}}\frac{\partial a^\ast}{\partial\ell^{(0)}_{\mathrm{inv}}}\,.
\eea
Thus the hidden relation between $C_T$ and $a^\ast$ manifests itself when the invariant length $\ell^{(0)}_{\mathrm{inv}}$ is chosen.  As we saw in the main text, more hidden relations exist between the central charge and OPE coefficients under this choice of invariant length.

\section{Stress-tensor three-point functions}
\label{app:3pt}

\subsection{Three-point structures}

Three-point function of stress-tensor $\langle TTT\rangle$ was first obtained in \cite{Erdmenger:1996yc}. The expression is quite cumbersome, we contract the three-point function with graviton polarizations (which are null $\epsilon^2=0$), where the indices can be recovered by using Toda operator \cite{Costa:2011mg}. We further fix the conformal frame $(0,x,\infty)$ to simplify the structures
\bea
\langle TTT\rangle &=& \fft{1}{2}\mathcal{A} x^{-d-4} \Bigg(\epsilon _2\cdot \epsilon _3 (2 x^2 \epsilon _1\cdot \epsilon _3 (x^2 \epsilon _1\cdot \epsilon _2-2 x\cdot \epsilon _1 x\cdot \epsilon _2)-(d^2-4) (x\cdot \epsilon _1)^2 x\cdot \epsilon _2 x\cdot \epsilon _3)
 \cr &&+(d^2-4) x\cdot \epsilon _1 x\cdot \epsilon _2 x\cdot \epsilon _3
 (\epsilon _1\cdot \epsilon _2 x\cdot \epsilon _3-\epsilon _1\cdot \epsilon _3 x\cdot \epsilon _2)\Bigg)\cr
&&+ \fft{1}{4} \mathcal{B} x^{-d-6} \Bigg(2 x^2 \epsilon _2\cdot \epsilon _3 x\cdot \epsilon _1 (x\cdot \epsilon _3 (-(d-2) x\cdot \epsilon _1 x\cdot \epsilon _2 -2 x^2 \epsilon _1\cdot \epsilon _2)+2 x^2 \epsilon _1\cdot \epsilon _3 x\cdot \epsilon _2)\cr
&&+x\cdot \epsilon _2 x\cdot \epsilon _3 \Big(x\cdot \epsilon _1 x\cdot \epsilon _3 (2 (d+2) x^2 \epsilon _1\cdot \epsilon _2+(d-2) (d+4) x\cdot \epsilon _1 x\cdot \epsilon _2)
 \cr &&-2 x^2 \epsilon _1\cdot \epsilon _3 ((d-2) x\cdot \epsilon _1 x\cdot \epsilon _2  +2 x^2 \epsilon _1\cdot \epsilon _2)\Big)\Bigg)\cr
&&+\frac{1}{2} \mathcal{C} x^{-d-6} \Bigg(4 x^2 x\cdot \epsilon _1 x\cdot \epsilon _2 x\cdot \epsilon _3 (d \epsilon _1\cdot \epsilon _3 x\cdot \epsilon _2-(d+2) \epsilon _1\cdot \epsilon _2 x\cdot \epsilon _3)\cr
&& +(x\cdot \epsilon _1)^2\Big(4 d x^2 \epsilon _2\cdot \epsilon _3 x\cdot \epsilon _2 x\cdot \epsilon _3 -(d-2) (d+4)(x\cdot \epsilon _2)^2 (x\cdot \epsilon _3)^2+2 x^4(\epsilon _2\cdot \epsilon _3)^2\Big)\cr
&&+2 x^4 \Big((\epsilon _1\cdot \epsilon _2)^2 (x\cdot \epsilon _3)^2 +(\epsilon _1\cdot \epsilon _3)^2(x\cdot \epsilon _2)^2\Big)\Bigg)\,.
\eea
In general, there are three independent structures with parameters $\mathcal{A}, \mathcal{B}$ and $\mathcal{C}$, which are interpreted as OPE coefficients of the associated OPE basis
\be
\langle TTT\rangle = (\mathcal{A},\mathcal{B},\mathcal{C})\cdot \langle TTT\rangle^{(i)}\,.
\ee
We can arbitrarily rotate the OPE basis, i.e., three-point structures $\langle TTT\rangle^{(i)}$, and read off the corresponding OPEs. One choice is the orthogonal basis \cite{Caron-Huot:2021kjy,Li:2021snj}
\be
\langle TTT\rangle=\sum_{i=1}^3\lambda_{TTT}^{(i)}\langle TTT\rangle^{(i)}_{\rm orth}\,,
\ee
where the orthogonality is defined with respect to three-point pairing \cite{Karateev:2018oml}. These are
\bea
&& \langle TTT\rangle^{(1)}_{\rm orth}=\ft{d (d^2-4) }{16 (d+1) (d+3)}x^{-d-6} \Bigg(4 x^2 x\cdot \epsilon _1 \Big((d+4) \epsilon _1\cdot \epsilon _3 x\cdot \epsilon _2-(d+2) \epsilon _1\cdot \epsilon _2 x\cdot \epsilon _3\Big)\cr
&& \times (d x\cdot \epsilon _2 x\cdot \epsilon _3 +2 x^2 \epsilon _2\cdot \epsilon _3)
+2 x^4 \Big(\epsilon _1\cdot \epsilon _3 ((d+4) \epsilon _1\cdot \epsilon _3 (x\cdot \epsilon _2)^2-4 x^2 \epsilon _1\cdot \epsilon _2 \epsilon _2\cdot \epsilon _3)\cr
&&+(d+4) (\epsilon _1\cdot \epsilon _2)^2 (x\cdot \epsilon _3)^2 -4 (d+2) \epsilon _1\cdot \epsilon _3 \epsilon _1\cdot \epsilon _2 x\cdot \epsilon _2 x\cdot \epsilon _3\Big)\cr
&&+(d+4) (x\cdot \epsilon _1)^2 (4 d x^2 \epsilon _2\cdot \epsilon _3 x\cdot \epsilon _2 x\cdot \epsilon _3+(d-2) d (x\cdot \epsilon _2)^2 (x\cdot \epsilon _3)^2+2 x^4 (\epsilon _2\cdot \epsilon _3)^2)\Bigg)\,,\cr
&& \langle TTT\rangle^{(2)}_{\rm orth}=-\ft{(d-2) d }{16 (d-1)^2}x^{-d-6}\Bigg(-2 x^4 \Big(\epsilon _1\cdot \epsilon _3 ((d^2+d-4) \epsilon _1\cdot \epsilon _3 (x\cdot \epsilon _2)^2 \cr
&&+4 (d+1) x^2 \epsilon _1\cdot \epsilon _2 \epsilon _2\cdot \epsilon _3) +(d^2+d-4) (\epsilon _1\cdot \epsilon _2)^2 (x\cdot \epsilon _3)^2-4 (d+1) \epsilon _1\cdot \epsilon _3 \epsilon _1\cdot \epsilon _2 x\cdot \epsilon _2 x\cdot \epsilon _3\Big)\cr
&&+(x\cdot \epsilon _1)^2 (-2 (d^2+d-4)x^4 (\epsilon _2\cdot \epsilon _3)^2
  \cr &&+4 (d-3) (d+2) x^2 \epsilon _2\cdot \epsilon _3 x\cdot \epsilon _2 x\cdot \epsilon _3+(d-3) (d-2) (d+2) (d+4) (x\cdot \epsilon _2)^2 (x\cdot \epsilon _3)^2)
  \cr &&+4 x^2 x\cdot \epsilon _1 (\epsilon _1\cdot \epsilon _2 x\cdot \epsilon _3+\epsilon _1\cdot \epsilon _3 x\cdot \epsilon _2) (2 (d+1) x^2 \epsilon _2\cdot \epsilon _3+(d-3) (d+2) x\cdot \epsilon _2 x\cdot \epsilon _3)\Bigg)\,,
\cr && \langle TTT\rangle^{(3)}_{\rm orth}=\ft{(d-2) d }{16 (d-1)}x^{-d-6}\Bigg(4 x^2 x\cdot \epsilon _1 \Big(d \epsilon _1\cdot \epsilon _2 x\cdot \epsilon _3-(d-2) \epsilon _1\cdot \epsilon _3 x\cdot \epsilon _2\Big)\cr
&&\times\Big((d+2) x\cdot \epsilon _2 x\cdot \epsilon _3-2 x^2 \epsilon _2\cdot \epsilon _3\Big) +2 x^4 \Big(\epsilon _1\cdot \epsilon _3 ((d-2) \epsilon _1\cdot \epsilon _3 (x\cdot \epsilon _2)^2+4 x^2 \epsilon _1\cdot \epsilon _2 \epsilon _2\cdot \epsilon _3)\cr
&&+(d-2) (\epsilon _1\cdot \epsilon _2)^2 (x\cdot \epsilon _3)^2-4 d \epsilon _1\cdot \epsilon _3 \epsilon _1\cdot \epsilon _2 x\cdot \epsilon _2 x\cdot \epsilon _3\Big) +(d-2) (x\cdot \epsilon _1)^2 \Big(\cr
&&-4 (d+2) x^2 \epsilon _2\cdot \epsilon _3 x\cdot \epsilon _2 x\cdot \epsilon _3+(d+2) (d+4) (x\cdot \epsilon _2)^2 (x\cdot \epsilon _3)^2+2 x^4 (\epsilon _2\cdot \epsilon _3)^2\Big)\Bigg)\,.
\eea
In the flat-space limit of $d=3$, this orthogonal basis reduces to three-point vertices $R$, $[R^{(2)}]$ (which is vanishing in $d=3$) and $[R^{(3)}]$. It is easy to show that $\langle TTT\rangle^{(2)}_{\rm orth}\equiv 0$ in $d=3$ by parameterizing $\epsilon=
(e_1,e_2,i\sqrt{e_1^2+e_2^2})$. It is then straightforward to find eq.~\eqref{OPE rotation}.

\subsection{OPEs in $d=4$}

Directly evaluating stress-tensor three-point function from Witten diagram is highly challenging because of unmanageable bulk tensor structures. Nevertheless, one can instead consider ANEC operator
\be
\mathcal{E}(x^{+},\vec{x}^{\bot})=\int dx^{-}T_{--}(x^+,x^-,\vec{x}^{\bot})\,,
\ee
where $x^{\pm}$ are lightcone directions. ANEC operator and more general light-ray operators \cite{Kologlu:2019mfz} are especially useful to constrain AdS gravity, e.g., bound $a/c$ and suggest EFT prescription \cite{Hofman:2008ar,Hofman:2016awc,Belin:2019mnx}, and can even sharply provide superconvergence sum rule \cite{Kologlu:2019bco}. To encode stress-tensor three-point function, we can evaluate the expectation value ANEC under states excited by the stress tensors \cite{Hofman:2008ar}
\be
\fft{\langle T\mathcal{E}T\rangle|_{x^+=0}}{\langle TT\rangle}\sim 1+t_2\big(\fft{(\epsilon_1\cdot n)(\epsilon_2\cdot n)}{\epsilon_1\cdot\epsilon_2}
-\fft{1}{d-1}\big)+t_4\big(\fft{(\epsilon_1\cdot n)^2(\epsilon_2\cdot n)^2}{(\epsilon_1\cdot\epsilon_2)^2}-\fft{2}{d^2-1}\big)\,,
\ee
where $\vec{x}^{\bot}=\vec{n}/(1+n^{d-1})$. The holographic side is much simplified, since the ANEC operator excites a shock wave in the bulk, namely
\be
ds^2=\fft{\ell_{\rm eff}^2}{\tilde{r}^2}d\tilde{r}^2+\tilde{r}^2(dx_idx^i+\delta(x^+)W(\rho,x)(dx^+)^2)\,,
\ee
where
\bea
W(\rho,x)\sim \tilde{r}^{-4} (|x_1-x_1^\prime|^2+|x_2-x_2|^\prime+\ell_{\rm eff}^2/\tilde{r}^2)^{-3}\,,
\eea
and we have made the following coordinate transformation from the Poincare AdS
\bea
x^+\rightarrow -\fft{1}{x^+}\,,\quad x^{-}\rightarrow x^{-}-\fft{|\vec{x}^{\bot}|^2}{x^+}-\fft{z^2}{x^{+}}\,,\quad \vec{x}^{\bot}\rightarrow
\fft{\vec{x}^\bot}{x^+}\,,\quad \tilde{r}= \fft{\ell_{\rm eff}x^+}{z}\,.
\eea
Consequently, we only have to find the contribution that the shock wave couples to two gravitons. After calculation, we find
\bea
t_2=-\frac{24}{C_T}(40g_{3,5}+8g_{3,6}+108g_{3,7}+45g_{3,8}-2\ell^2_{\mathrm{eff}}g_{2,3}
)\,,\quad
t_4=\frac{2160}{C_T}\left(2g_{3,7}+g_{3,8}\right)\,.
\eea
(The generalize the result was obtained in \cite{Li:2019auk}.) We can readily verify that they are field redefinition invariant. Use the identities \cite{Hofman:2008ar}
\be
t_2=\frac{15 (5 \mathcal{A}+4 \mathcal{B}-12 \mathcal{C})}{9 \mathcal{A}-\mathcal{B}-10 \mathcal{C}}\,,\qquad t_4=-\frac{15 (17 \mathcal{A}+32 \mathcal{B}-80 \mathcal{C})}{4 (9 \mathcal{A}-\mathcal{B}-10\mathcal{C})}\,,\quad C_T=\fft{\pi^2}{12}(9\mathcal{A}-\mathcal{B}-10\mathcal{C})\,,\label{t2-t4-ABC}
\ee
we find, in terms of $\ell_{\rm inv}^{(0)}$, that
\bea
\mathcal{A}&=&-\fft{64 (\ell_{\rm inv}^{(0)})^3}{9 \pi^5 G_N^{(5)}}\Big(1- \fft{24q_3}{M^2(\ell_{\rm inv}^{(0)})^2}+\fft{6}{7M^4 (\ell_{\rm inv}^{(0)})^4}\big(
32(2c_7+c_8) -21 \hat q_3\big)\Big)\,,\cr
\mathcal{B}& =&-\fft{196(\ell_{\rm inv}^{(0)})^3}{9 \pi^5 G_N^{(5)}}\Big(1- \fft{636q_3}{49M^2(\ell_{\rm inv}^{(0)})^2}+\fft{6}{343M^4(\ell_{\rm inv}^{(0)})^4} \big(2918(2c_7+ c_8)-399 \hat {q}_3\big)\Big)\,,
\cr &&
\cr  \mathcal{C}&=&-\fft{92(\ell_{\rm inv}^{(0)})^3}{9 \pi^5 G_N^{(5)}}\Big(1-\fft{336q_3}{23M^2(\ell_{\rm inv}^{(0)})^2}+\fft{6}{161M^4(\ell_{\rm inv}^{(0)})^4} \big(142 (2c_7+c_8)-231 \hat{q}_3\big)\Big)\,.
\eea
where
\be
\hat q_3=\tilde{q}_3+\frac{8}{3}(5q_1+q_2-q_3)q_3-20 c_5 - 4 c_6 - \fft{30}{7} c_7 +
\fft{33}{14} c_8\,,
\ee
is an invariant combination under the field redefinition. Before proceeding further, we would like to recall that $\ell_{\rm inv}^{(0)}$ is the invariant expression associated with the effective AdS radius of massless gravities, satisfying the massless conditions $q_1=q_3, q_2=-4q_3$ and $\tilde q_3=0$, (the Gauss-Bonnet combination), together with \cite{Li:2017txk}
\bea
c_1 &=&\fft1{200} (16 c_3 + 6 c_4 + 160 c_5 + 40 c_6 + 52 c_7 - 9 c_8)\,,\cr
 c_2&=&\fft1{20} (-12 c_3 - 7 c_4 - 80 c_5 - 20 c_6 - 24 c_7 + 3 c_8)\,.
\eea
There are six-parameter family of cubic massless gravities, but the $(a,c)$ charges depend only on $(c_5,c_6,c_7,c_8)$ \cite{Li:2017txk}.  The same is true for the OPE coefficients $({\mathcal A}, {\mathcal B}, {\mathcal C})$. In fact they are grouped into two combinations $(2c_7+ c_8)$ and $\hat q_3$. From the perturbative perspective, the field redefinition can be used to reduced the family of theories further down to one cubic theory, e.g.~with a non-vanishing $c_8$, since the cubic Lovelock combination with $2c_7+c_8=0$ vanishes identically in $D=5$ bulk dimensions. We can thus consider a fiducial $\ell_{\rm inv}^{(0)}$ associated with quasi-topological gravity $(c_5=3/8, c_6=-9/7, c_7=0, c_8=1)$, for which $\hat q_3$ vanishes.
Since the central charges can be expressed as linear combinations of the OPE coefficients.  It is instructive to define the following linearly independent combinations
\bea
\tilde {\mathcal{A}} &\equiv& \frac{\pi ^6 }{2880}(13 \mathcal{A}-2 \mathcal{B}-40 \mathcal{C})=a\,,\cr
\tilde {\mathcal{B}} &\equiv& \fft{\pi^6}{2016}(9 \mathcal{A}-2 \mathcal{B}-2\mathcal{C})\,,\cr
\tilde {\mathcal{C}} &\equiv&  \frac{\pi ^6 }{480} (9 \mathcal{A}-\mathcal{B} -10 \mathcal{C})=c\,.
\eea
There are hidden linear differential relations among these OPE coefficients, given by eq.~\eqref{hidden}.  The above choice of the linear combinations for ${\mathcal{A}}$ and ${\mathcal{C}}$ is unique, but it is not for the $\tilde {\mathcal{B}}$.  If we instead choose
$\hat q_3 = z(2c_7+c_8)$, we can define
\be
\tilde {\mathcal{B}}=
\fft{\pi^6}{362880}\Big((1620 - 119 z)\mathcal{A}- 8(45 +28 z)\mathcal{B}-40(9 - 14 z)\mathcal{C}\Big)\,,
\ee
which also satisfies \eqref{hidden}.  Since $(\tilde {\mathcal{A}},\tilde {\mathcal{B}},\tilde {\mathcal{C}})$ are linearly independent, different choices of $z$ specifies different basis for the OPE coefficients, but the differential relations \eqref{hidden} is unchanged. Thus the $(a,c)$ relation \eqref{relation 4d} is valid for all massless gravity even in the non-perturbative sense where the coupling constants are not necessarily small, the statements in \eqref{hidden} is valid only perturbatively where the OPE coefficients can be defined.

\end{document}